\documentclass[pre]{revtex4}
\usepackage[dvips]{graphicx}

\begin{document}
\title{Local prediction of turning points of oscillating time series}
\author{D. Kugiumtzis}
 \email{dkugiu@gen.auth.gr}
 \homepage{http://users.auth.gr/dkugiu}
 \affiliation{Department of Mathematical, Physical and Computational Sciences, Faculty of
Engineering, Aristotle University of Thessaloniki, Thessaloniki 54124, Greece }

\begin{abstract}
For oscillating time series, the prediction is often focused on the
turning points. In order to predict the turning point magnitudes and
times it is proposed to form the state space reconstruction only
from the turning points and modify the local (nearest neighbor)
model accordingly. The model on turning points gives optimal
prediction at a lower dimensional state space than the optimal local
model applied directly on the oscillating time series and is thus
computationally more efficient. Monte Carlo simulations on different
oscillating nonlinear systems showed that it gives better
predictions of turning points and this is confirmed also for the
time series of annual sunspots and total stress in a plastic
deformation experiment.
\end{abstract}


\maketitle

The prediction of oscillating time series that do not exhibit
apparent periodicity has been a long lasting challenge and the focus
of three time series prediction competitions
\cite{Weigend94,Suykens98,ESTSP07}. Evidence from the winning models
of the competitions and other prediction studies raises two main
points: (a) multi-step ahead prediction requires a long time window
and consequently a high embedding dimension $M$, and (b) local
prediction models, called also nearest neighbor models, are
computationally efficient and compete other more complicated
black-box models, such as neural networks. Notably, they were among the
winning entries of the two first competitions. In this work these
two points are incorporated in the prediction of the turning points
of the oscillating time series. The prediction of successive samples
is associated with intra-oscillation correlations whereas the
prediction of turning points regards inter-oscillation correlations,
which are more relevant to the underlying oscillating dynamics
\cite{Kugiumtzis04a}. Turning point prediction is of great practical
interest in many applications, such as finance
\cite{Garcia98,Bao08}. It will be shown below that the prediction of
turning points with local models can be improved using state space
reconstruction solely on the turning points at a lower embedding
dimension $m$.

For an oscillating time series of length $N$,
$\{x(t)\}_{t=1}^N$, where the observation time is $t\tau_s$ and
$\tau_s$ is the sampling time, a sample $x(t)$ is a turning
point if it is the minimum or maximum of all samples in the
time window $[t-p,t+p]$ where the parameter $p$ determines the
tolerance for temporal closeness of successive turning points.
A small $p$ may assign turning points for glitches in the case of noisy
oscillations, whereas a large $p$ may not detect peaks and
troughs of short lasted oscillations. For noisy time series,
a small $p$ can still be used in conjunction with filtering, and then
the turning points are located on the smoothed time series. Note that the
turning point magnitudes are then taken from the original time series.
Denoting the turning point $y_i=x(t_i)$ at time point $t_i$, we derive the time
series $\{y_i\}_{i=1}^n$ and $\{t_i\}_{i=1}^n$ of magnitudes
and times of the alternating turning points, respectively. Thus
two successive samples of $\{y_i\}_{i=1}^n$ regard an
oscillation of $\{x(t)\}_{t=1}^N$. One may also consider three
turning points to include both the start and end of the
oscillation.

In the reconstruction of the $M$-dimensional pseudo-state space
from $\{x(t)\}_{t=1}^N$, the reconstructed points have the
general form
$\mathbf{x}(t)=[x(t),x(t-\tau_1),\ldots,x(t-\tau_{M-1})]^{\prime}$.
The standard delay embedding suggests the use of a properly
selected fixed delay $\tau$, so that the time window length is
$\tau_w=\tau_{M-1}=(M-1)\tau$. For multi-step prediction, $\tau_w$ should be large enough to account for the mean
orbital period of the underlying to time series trajectory
and it should cover the period of an oscillation or a pattern of
oscillations \cite{Kugiumtzis96}. The choice of large delays
or a fixed large $\tau$ instead of a large $M$ would not be
appropriate as in this case large pieces of information from
the oscillation, most importantly the peak and trough, may not
be represented in $\mathbf{x}(t)$.
The main idea in the proposed approach is to let $\tau_j$,
$j=1,\ldots,M-1$, and $\tau_w$ vary with the target time $t$,
so that the peaks and troughs are selected as components of
$\mathbf{x}(t)$, resulting in a smaller embedding dimension $m$
than $M$. Moreover, the reconstructed trajectory
$\{\mathbf{x}(t)\}_{1+\tau_w}^N$ is subsampled at times
$\{t_i\}_{i=1}^n$. This is actually the state space
reconstruction of $\{y_i\}_{i=1}^n$ in an $m$-dimensional state
space. The reconstructed point is
\begin{equation}
 \begin{array}{rcl}
 \mathbf{y}_i & = & [y_i,y_{i-1},\ldots,y_{i-m+1}]^{\prime} \\
 & = & [x(t_i),x(t_{i-1}),\ldots,x(t_{i-m+1})]^{\prime},
 \end{array}
 \label{eq:embedextpoi}
\end{equation}
for $i=m,\ldots,n$, where the implied lags with regard to
$\mathbf{x}(t)$ are $\tau_j=t_i - t_{i-j}$, $j=1,\ldots,m-1$.

\begin{figure}[h!]
\centerline{\hbox{\includegraphics[height=60mm]{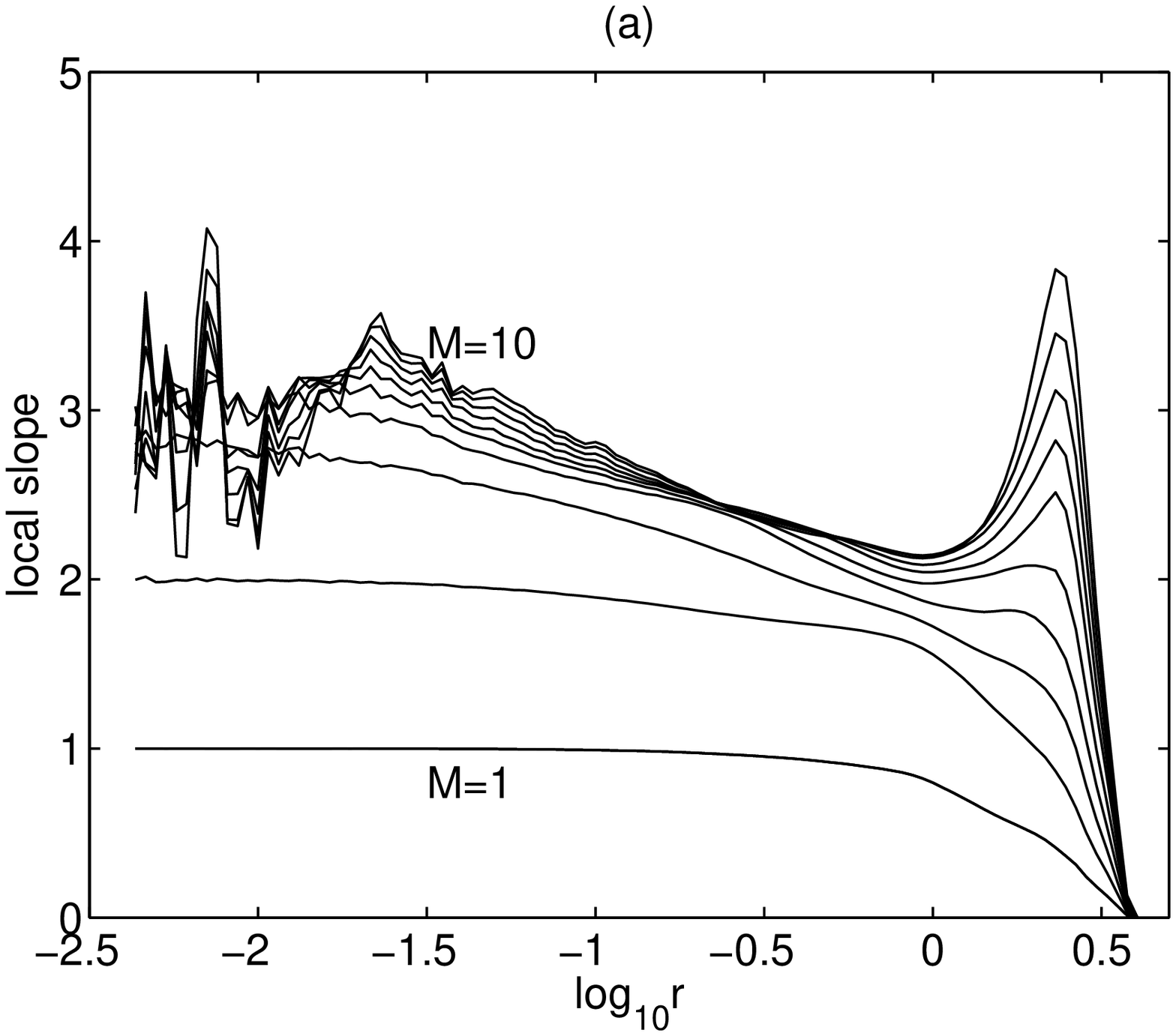}
 \includegraphics[height=60mm]{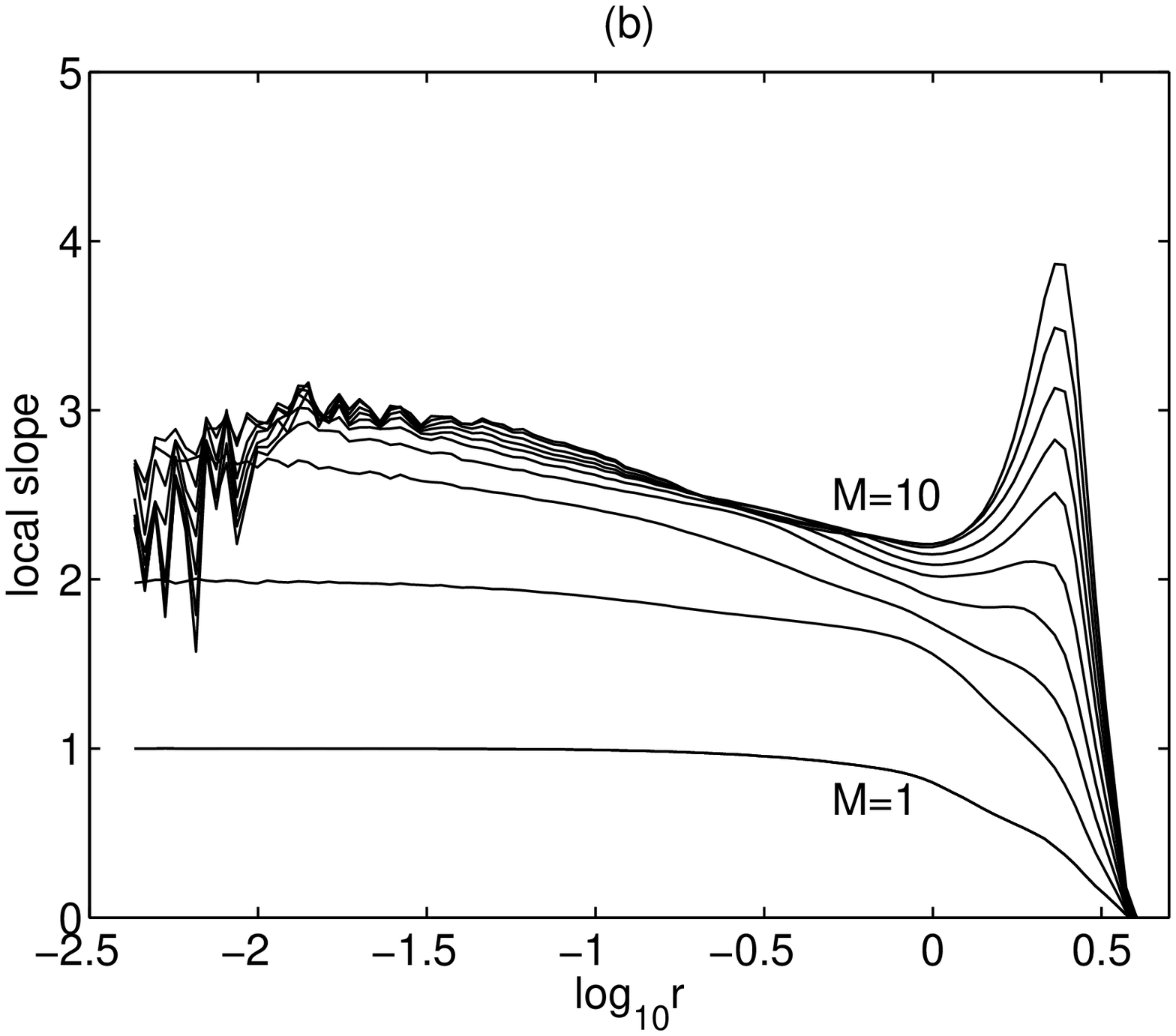}}}
 \centerline{\hbox{\includegraphics[height=60mm]{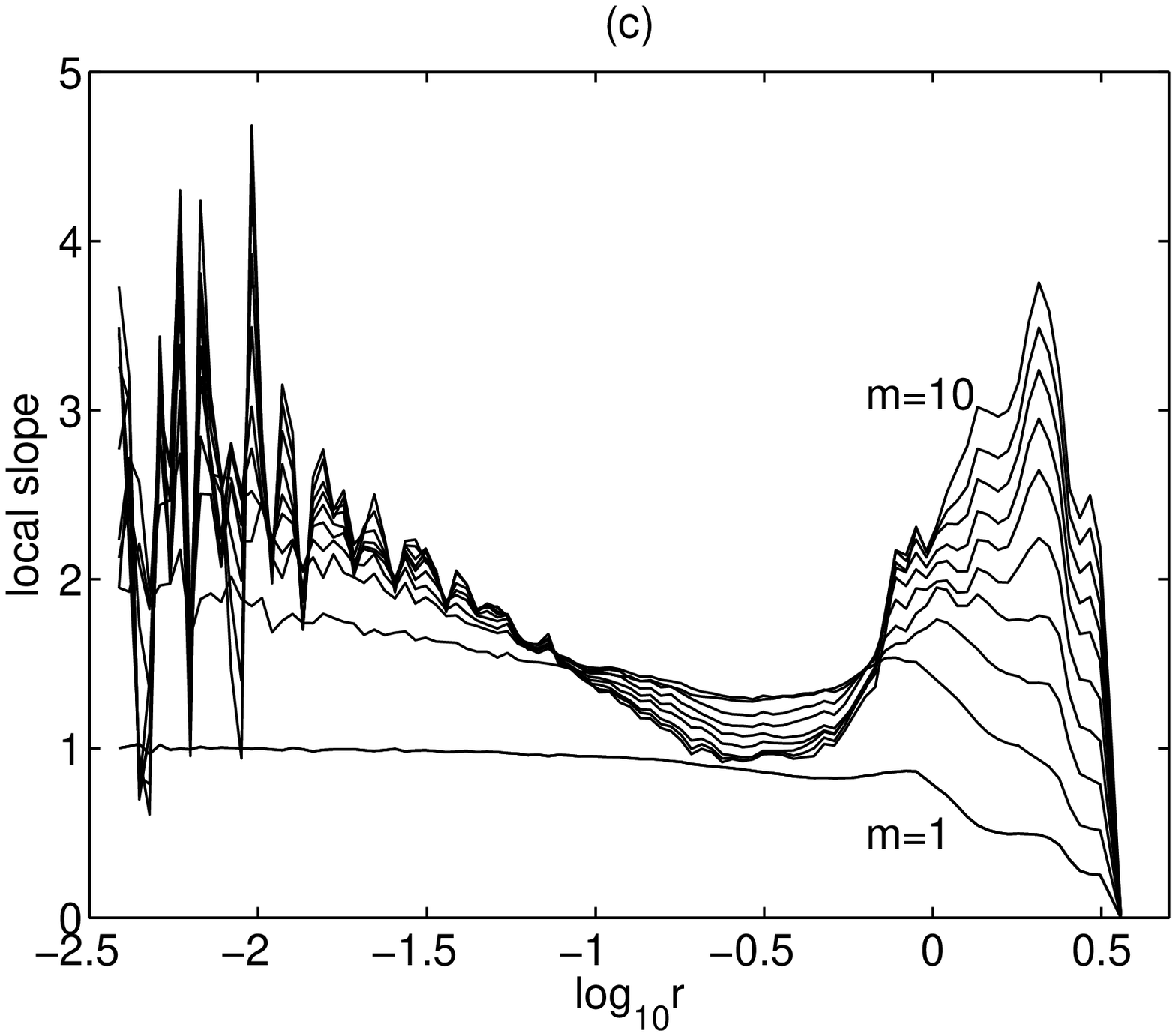}
 \includegraphics[height=60mm]{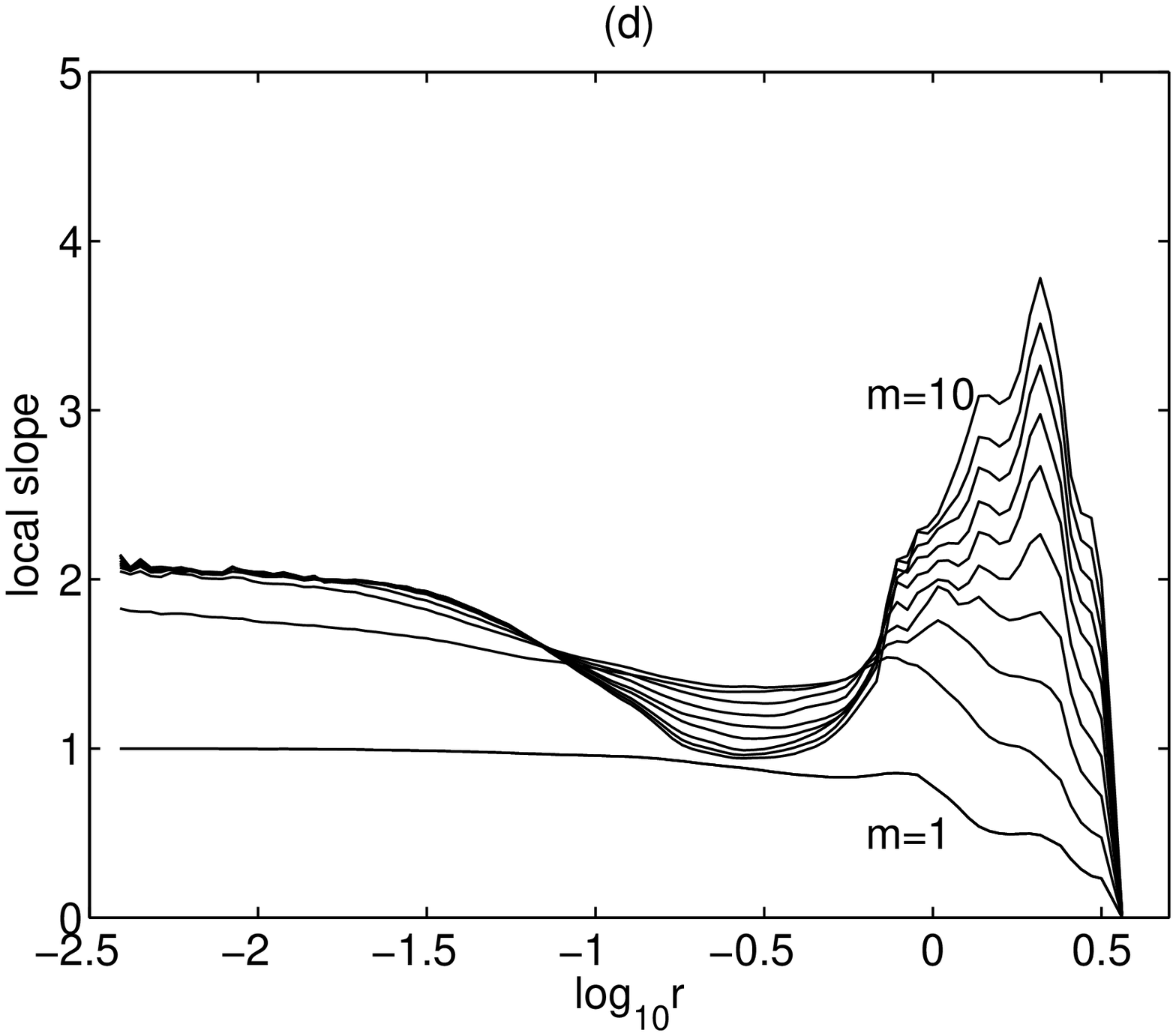}}}
\caption{Local slope vs base 10 logarithm of distance $r$ for embedding
dimensions $1,\ldots,10$, as indicated in the panels, and for
oscillating time series in (a) and (b) and turning point time
series in (c) and (d) from the Mackey Glass system with
$\Delta=30$. (a) $\tau_s=5\mbox{s}$, $N=200000$, $\tau=35$; (b)
$\tau_s=30\mbox{s}$, $N=200000$, $\tau=6$; (c)
$\tau_s=5\mbox{s}$, $N=200000$, $n=6380$; (d)
$\tau_s=5\mbox{s}$, $N=6284867$, $n=200000$. The lag $\tau$ is
selected from the minimum of mutual information.}
 \label{fig:corext}
\end{figure}

The compression of $\{x(t)\}_{t=1}^N$ to $\{y_i\}_{i=1}^n$
simplifies the embedding because $m<M$ and no lag parameter is
involved, at the cost of stripping off the information in the
samples between the turning points. This compression is
analogue to the reduction of a flow to its Poincar\'{e} map.
Note that formally Poincare maps require that a state space
reconstruction is made first, whereas in this approach the
state space reconstruction is made on the turning points.
For low-dimensional chaotic systems with sheet-like structure,
it has been shown that the local maxima, i.e. every second
turning point, reproduce the dynamics of the respective
Poincar\'{e} map that has fractal dimension one less than that
of the flow, e.g. the time series of the third variable of the
Lorenz system \cite{Ott94} (see also \cite{Candaten00}
for the so-called peak-to-peak dynamics). Similarly, we expect
that the time series of magnitudes of turning points preserve
the original dynamics of the flow and form an attractor with a fractal
dimension smaller than that for the flow and somehow larger
than that for the corresponding Poincar\'{e} map, as estimated
through the local maxima. This is demonstrated in
Fig.~\ref{fig:corext} for the Mackey Glass delay differential
equation with delay $\Delta=30$ that regards a correlation
dimension $\nu \simeq 3.0$ \cite{Mackey77}. The delay differential
equation is solved with a discretization step of 0.1s.
The local slope of the correlation integral estimated on a
densely sampled oscillating time series, i.e. $\tau_s=5\,\mbox{s}$,
$N=200000$ and roughly 3200 oscillations, does not maintain
sufficient scaling and the same holds for the respective time
series of turning points (see Fig.~\ref{fig:corext}a and c).
The insufficient scaling persists for the oscillating time
series even for a larger sampling time ($\tau_s=30\,\mbox{s}$) giving 6 times more
oscillations for the same $N$ (see Fig.~\ref{fig:corext}b),
whereas scaling at the level $\nu \simeq 2.2$ is formed from
the turning point time series of the same $N$, as shown in
Fig.~\ref{fig:corext}d.

The amount of loss of information in the compression of
$\{x(t)\}_{t=1}^N$ to $\{y_i\}_{i=1}^n$ and $\{t_i\}_{i=1}^n$
depends on the curvature of the upward and downward pattern of
the oscillations. Actually, in the case of linear upward and
downward trends there is no loss of
information, as each sample $x(t_i-k)$ between two turning
points $x(t_{i-1})$ and $x(t_i)$, where $k \in
\{0,1,\ldots,t_i-t_{i-1}\}$, can be expressed in terms of the
magnitude and time of the two turning points as
\[
x(t_i-k) = x(t_i)-k\frac{x(t_i)-x(t_{i-1})}{t_i-t_{i-1}}=
 y_i-k\frac{y_i-y_{i-1}}{t_i-t_{i-1}}.
\]
The delay embedding on $\{y_i\}_{i=1}^n$ does not account for the
times $\{t_i\}_{i=1}^n$ of the turning points and information from
the samples is lost. However, for prediction purposes, it is
operationally tractable to form the reconstructed state space from
the magnitudes $\{y_i\}_{i=1}^n$ in order to find neighboring points
for the local prediction scheme and then call in the times
$\{t_i\}_{i=1}^n$ to estimate the time position that corresponds to
the predicted magnitude of the turning point, as shown below. Our
attempts on simulated chaotic systems with dynamic local regression
models making use of magnitudes and times to reconstruct the state
space showed no improvement in the prediction of turning points.

The prediction model of choice in this work is the local
average mapping (LAM), but other local models can be developed
in a similar way. For a fixed number of neighbors $K$ and given
the turning points up to time $t_i$, the one-step ahead
prediction of the turning point magnitude $y_{i+1}$ is
estimated by the average of the one-step ahead mappings of the
$K$ nearest neighboring points $\mathbf{y}_{i(k)}$,
$k=1,\ldots,K$, to the target point $\mathbf{y}_i$
\begin{equation}
\hat{y}_{i+1} = \frac{1}{K}\sum_{k=1}^K y_{i(k)+1}.
 \label{eq:extpoiT1}
\end{equation}
The prediction of the time of the turning point
$y_{i+1}$, $t_{i+1}$, is estimated from the average of the
corresponding time increments of the $K$ neighboring points
\begin{equation}
\hat{t}_{i+1} = t_i + \frac{1}{K}\sum_{k=1}^K(t_{i(k)+1}-t_{i(k)}).
 \label{eq:exttimT1}
\end{equation}

For the iterative prediction of the turning point magnitude at a
lead time $T$, the target point at time $t_{i+1}$ is updated as
$\mathbf{y}_{i+1}=[\hat{y}_{i+1},y_i,\ldots,y_{i-m+2}]^{\prime}$,
and the one-step prediction is done as in (\ref{eq:extpoiT1}) but
for the new set of neighboring points of $\mathbf{y}_{i+1}$. This
step is repeated until the prediction of $y_{i+T}$ is reached. The
direct prediction scheme is simpler and faster as it predicts
$y_{i+T}$ directly by the average of the $T$-step ahead mappings of
the neighboring points of $\mathbf{y}_i$. The iterative and direct
multi-step ahead prediction of the times of the turning points is
done similarly using (\ref{eq:exttimT1}). We refer to this
prediction model as {\em extreme magnitude local average map}
(EMLAM) to stress that the neighboring points in the local average
map are formed only on the basis of the magnitudes of the local
extremes (turning points).

\begin{figure}[h!]
\centering
\centerline{\includegraphics[height=60mm]{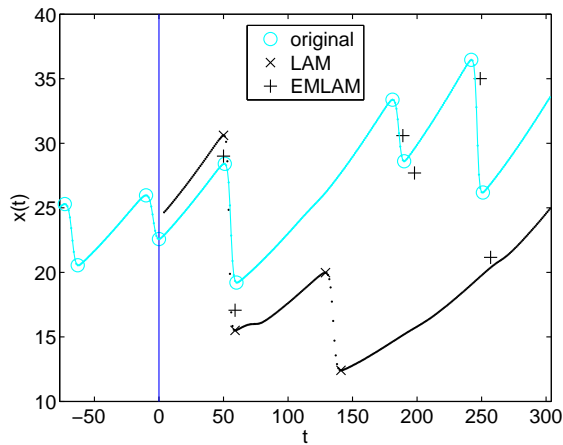}}
\caption{Out-of-sample direct predictions of turning points for the
fourth variable of the R\"{o}ssler hyperchaos system
($\tau_s=0.1\mbox{s}$) with LAM ($M=9$, $\tau=10$, $K=5$) and EMLAM
($m=3$, $K=5$) as given in the legend. The vertical line is at
$t=15000$ of the current turning point set to 0 for clarity. For LAM
$t$ is advanced by $p=3$ to account for the time the target turning
point is detected.}
 \label{fig:multistepprediction}
\end{figure}

We investigate whether we can predict the forthcoming local
extremes of an oscillating time series better than standard
multi-step prediction with LAM. In
Fig.~\ref{fig:multistepprediction} an example is shown for
multi-step prediction of turning points of the fourth variable
of the R\"{o}ssler hyperchaos system \cite{Roessler79}.
The predicted turning points with LAM, are identified from the multi-step 
sample predictions in the same way as the turning points are determined 
on the oscillating time series. This time series has a
rather linear upward and downward trend, so that the loss of
information using only the turning point time series is
expected to be small.

\begin{figure}[h!]
\centering
\centerline{\hbox{\includegraphics[height=60mm]{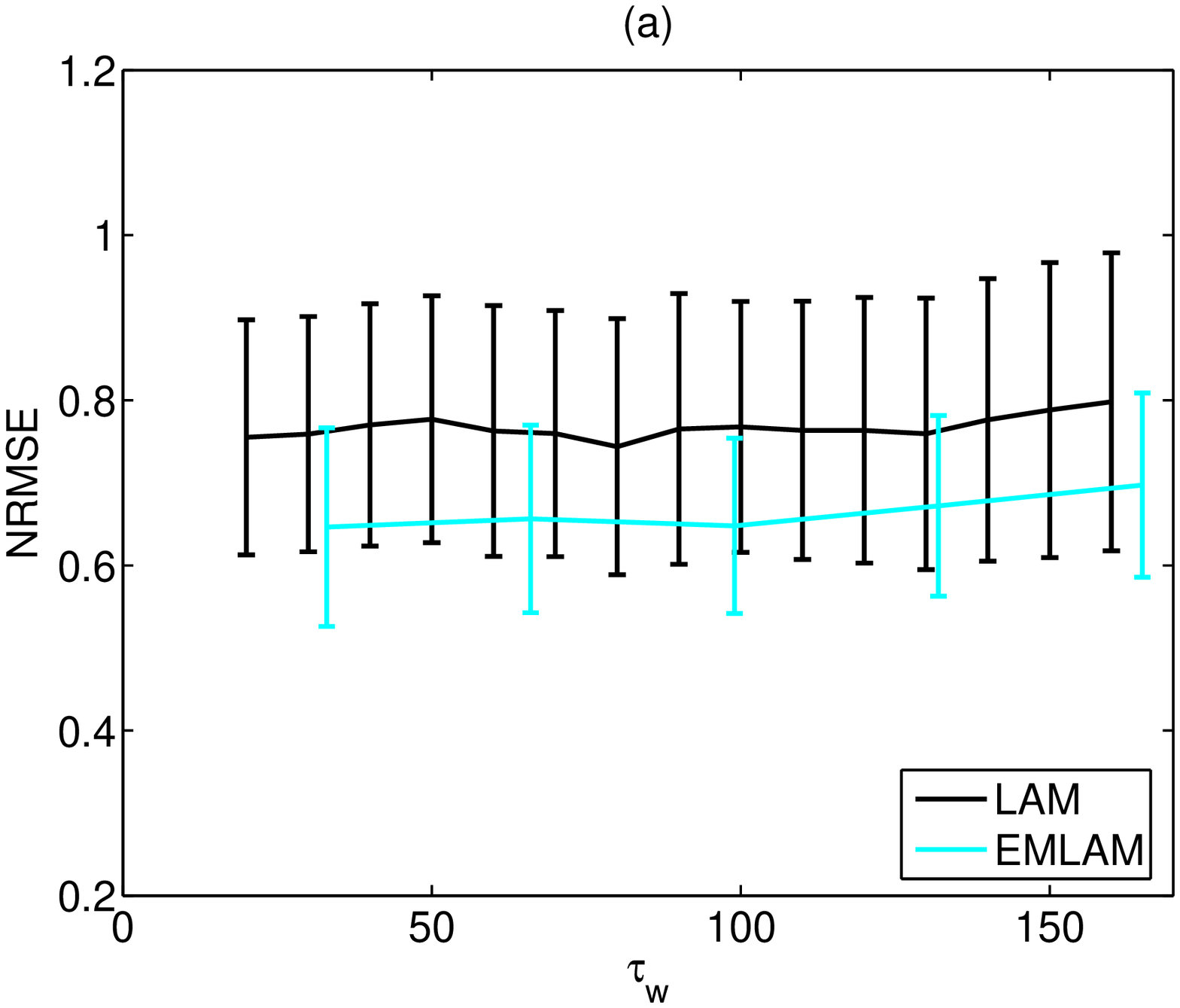}
\includegraphics[height=60mm]{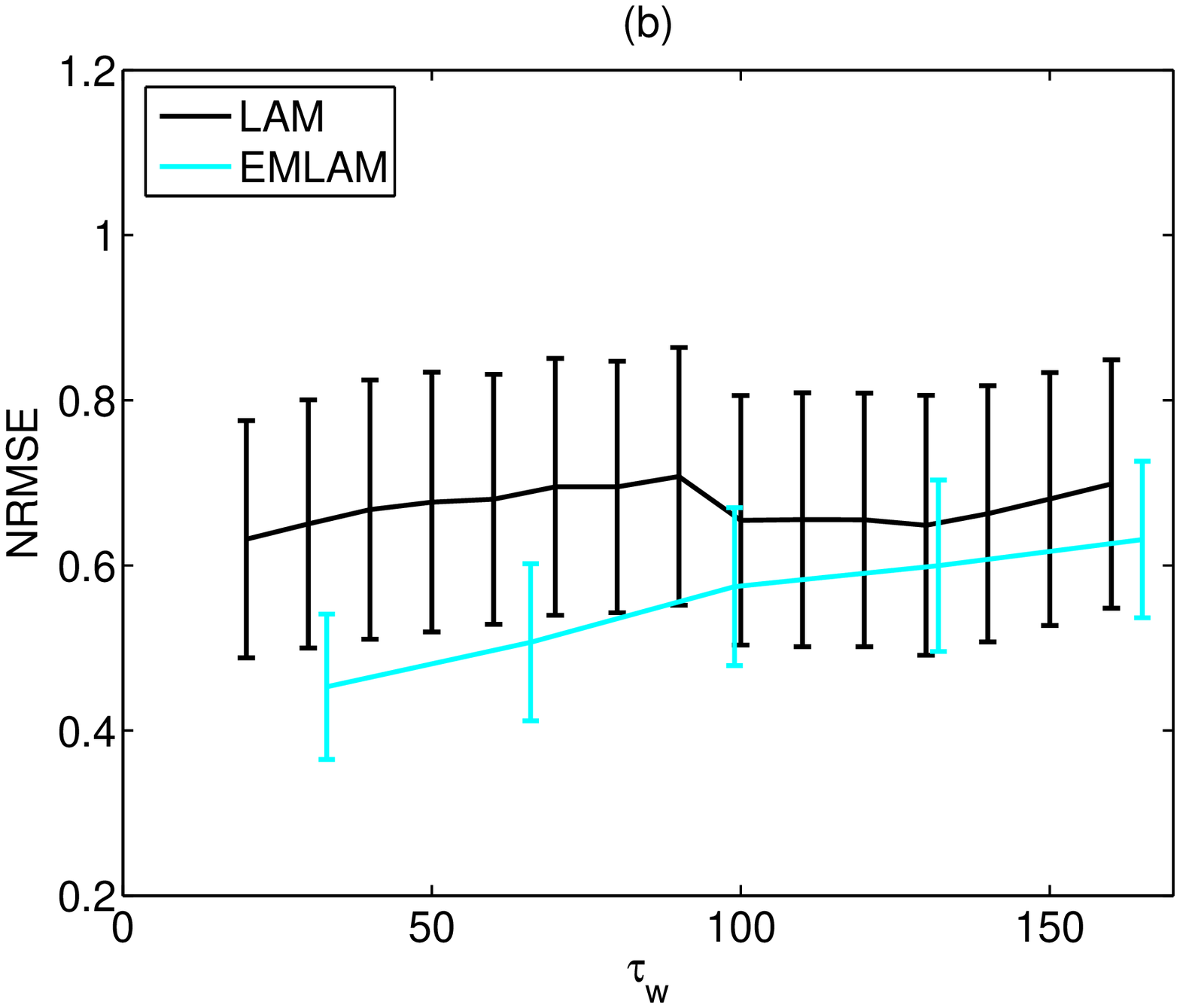}}}
\centerline{\hbox{\includegraphics[height=60mm]{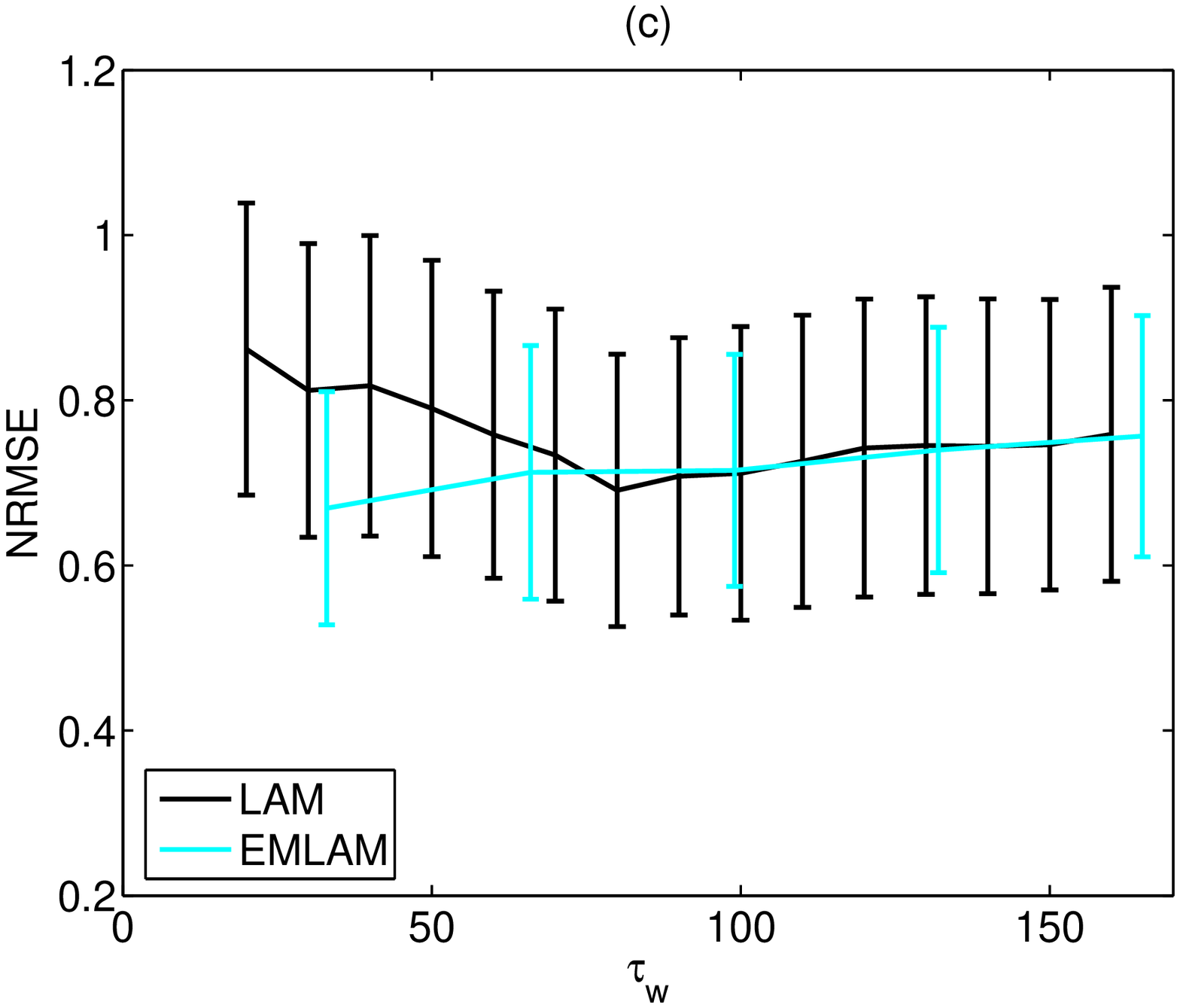}
\includegraphics[height=60mm]{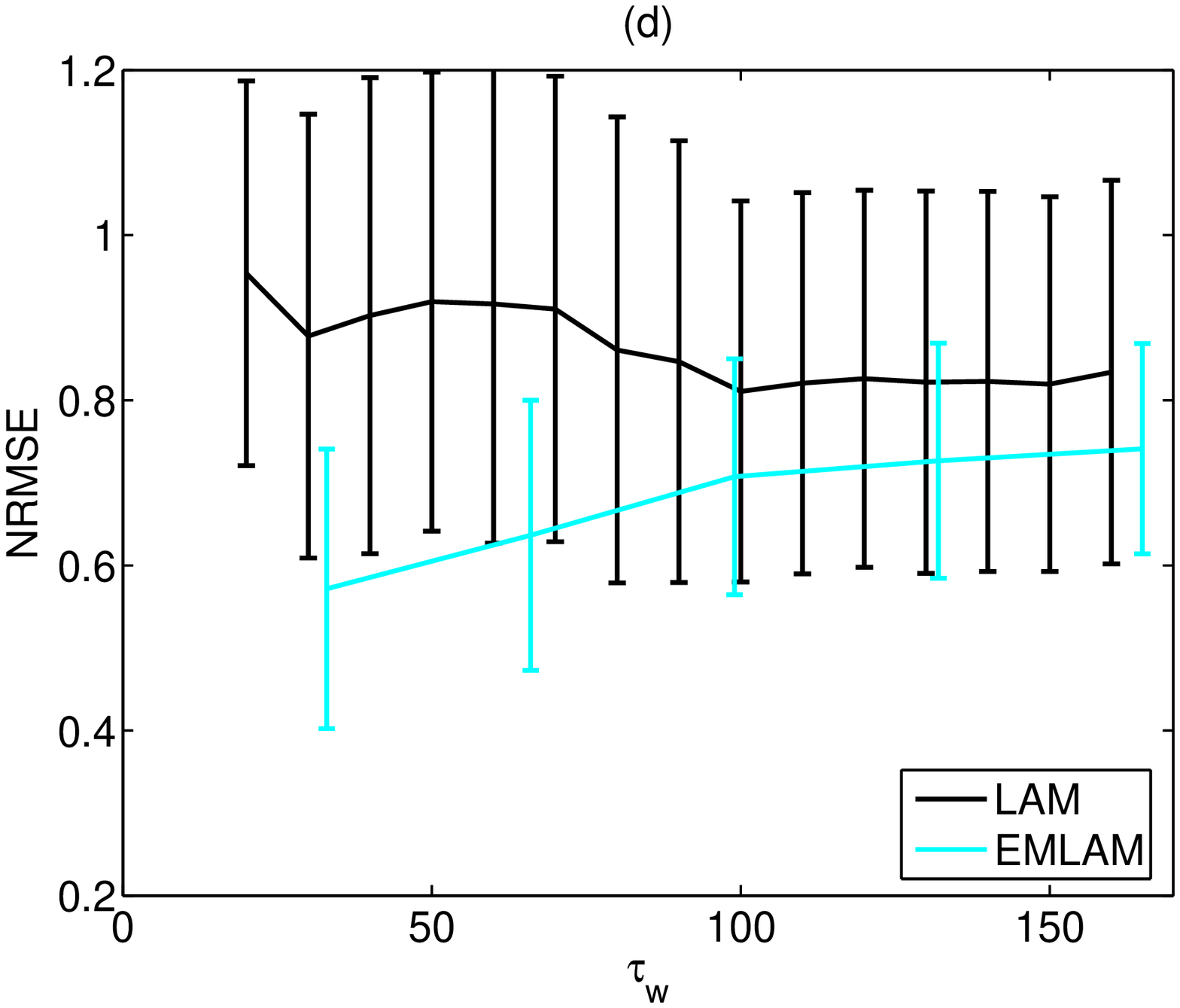}}}
\caption{The average NRMSE of the prediction of next turning point with LAM (direct
scheme) and EMLAM ($K=5$) from 1000 time series of length 8192 from
the fourth variable of the R\"{o}ssler hyperchaos system. The error bars 
denote the standard deviation of NRMSE. The
$\tau_w$ in the abscissa is defined as $\tau_w=(M-1)10$ for LAM and
$\tau_w=(m-1)33$ for EMLAM, as the mean oscillation period is
estimated from the power spectrum peak to be 66. In (a) and (b) the
data are noise-free and the prediction is for the magnitude and time
of the turning point, respectively. In (c) and (d) the same
predictions are for data corrupted with 10\% observational white
normal noise. In order to detect the turning points a zero-phase
filtering of order 13 is applied to the noisy data.}
 \label{fig:MCHyperChaos}
\end{figure}

The superiority of EMLAM over LAM for this system is confirmed
by simulations on 1000 realizations using the prediction measure
of normalized root mean square error (NRMSE) on the last quarter
of each time series. As shown in Fig.~\ref{fig:MCHyperChaos},
the prediction with EMLAM is better both for the magnitude
and time of the next turning point and this holds for noise-free
and noisy data. The difference is smaller for the turning point magnitudes
of the noisy data.
The difference of LAM and EMLAM prediction is larger when the
iterative scheme is used for predictions with LAM.
The superiority of EMLAM over LAM prediction persists for
different data sizes $N$, number of nearest neighbors $K$, and
prediction steps $T$, as shown in Table~\ref{tab:MCHyperChaos}.
Table~\ref{tab:MCHyperChaos} shows the least NRMSE and the
corresponding embedding dimension for the range of the other
factors. These are $K=1,5,10$, $T=1,2,3$, $N=4096,16384$, and 
are shown in the rows, whereas magnitude and time are in the columns.

\begin{table}[h!]
  \centering
   \begin{tabular}{|c|c|cc|cc|cc|cc|}
    \hline
    \multicolumn{2}{|c|}{} & \multicolumn{4}{c|}{Magnitudes} & \multicolumn{4}{c|}{Times}
    \\ \hline
    $T$ & $K$ & $M$ & \hspace{3mm} LAM & $m$ & EMLAM & $M$ & \hspace{3mm} LAM & $m$ & EMLAM \\ \hline
    & & \multicolumn{8}{|c|}{$N=4096$} \\ \hline
 1 & 1  & 10 & 1.144  & 3 & 0.873  & 9 & 1.083  & 2 & 0.556 \\
 1 & 5  & 10 & 1.318  & 4 & 0.816  & 10 & 1.247  & 2 & 0.562 \\
 1 & 10  & 10 & 1.479  & 4 & 0.885  & 10 & 1.336  & 2 & 0.645 \\
\hline
 2 & 1  & 9 & 1.137  & 3 & 1.013  & 8 & 1.569  & 2 & 1.097 \\
 2 & 5  & 10 & 1.259  & 4 & 0.916  & 7 & 1.710  & 2 & 0.976 \\
 2 & 10  & 10 & 1.396  & 4 & 0.935  & 8 & 1.769  & 2 & 0.998 \\
\hline
 3 & 1  & 10 & 1.352  & 3 & 1.153  & 8 & 2.245  & 2 & 1.438 \\
 3 & 5  & 10 & 1.673  & 4 & 1.028  & 7 & 2.450  & 5 & 1.224 \\
 3 & 10  & 10 & 1.940  & 2 & 1.005  & 8 & 2.577  & 4 & 1.213 \\
\hline
    & & \multicolumn{8}{|c|}{$N=16384$} \\ \hline
 1 & 1  & 10 & 0.813  & 3 & 0.508  & 9 & 0.911  & 2 & 0.369 \\
 1 & 5  & 10 & 0.928  & 2 & 0.543  & 10 & 1.179  & 2 & 0.374 \\
 1 & 10  & 10 & 1.150  & 4 & 0.582  & 10 & 1.331  & 2 & 0.426 \\
\hline
 2 & 1  & 9 & 0.837  & 3 & 0.672  & 8 & 1.290  & 3 & 0.801 \\
 2 & 5  & 10 & 0.956  & 3 & 0.678  & 9 & 1.649  & 3 & 0.783 \\
 2 & 10  & 10 & 1.157  & 4 & 0.681  & 10 & 1.737  & 3 & 0.808 \\
\hline
 3 & 1  & 9 & 1.013  & 3 & 0.794  & 8 & 1.929  & 3 & 1.136 \\
 3 & 5  & 4 & 1.411  & 4 & 0.787  & 4 & 2.460  & 3 & 1.085 \\
 3 & 10  & 10 & 1.547  & 4 & 0.767  & 3 & 2.476  & 5 & 1.084 \\
\hline
  \end{tabular}
  \caption{Summary results of the average NRMSE as for the noise-free case in Fig.~\ref{fig:MCHyperChaos} but for
varying $N$, $T$ and $K$. For each combination of $N$, $T$ and
$K$, the $M$ of best prediction with LAM and $m$ of best
prediction with EMLAM together with the respective NRMSE are
given, where $M=3,\ldots,10$ ($\tau=10$) and $m=2,\ldots,6$.}
  \label{tab:MCHyperChaos}
\end{table}
EMLAM provides computationally efficient predictions at a small
embedding dimension up to $m=4$, whereas LAM fails, at cases
dramatically, to reach the level of prediction of EMLAM for any of
the tested embedding dimensions. The direct prediction scheme shows
less dramatic differences in the performance of the two prediction
models. With the addition of observational noise the results are
qualitatively the same and the differences get smaller for the
direct prediction scheme (see also Fig.3) and larger for the
iterative prediction scheme. These results are based on simulations with 5\% and
10\% observational noise, not shown here. For the detection of the noisy
turning points, zero-phase filtering was used with an order adjusted
to the amount of noise in the data, in order to smooth out close
local maxima and minima that apparently do not correspond to real
oscillations.

The same simulations have been applied to other
oscillating time series of varying complexity that are not
characterized by linear upward and downward trend, namely the
first variable of the R\"{o}ssler hyperchaos system, the first
and third variable of the R\"{o}ssler system \cite{Roessler76},
and the Mackey-Glass delay differential equation for delay
parameter 17, 30, and 100 \cite{Mackey77}. The overall results
show that EMLAM gives as good, and at cases better, predictions
of turning points as the ones obtained by LAM.

The simulations revealed some important features of turning
point prediction in favor of EMLAM. In all cases, the best
predictions with EMLAM were obtained with a small $m$ at the
level of the fractal dimension of the underlying system, e.g.
for the Mackey-Glass system with delay 100 that has a fractal
dimension about 7, best results were obtained for $m$ at the
range from 7 to 10. For LAM, best results could be reached
only for large $\tau_w$ implying very large $M$. Another
interesting feature is that for LAM the direct scheme predicts
the turning points better than the iterative scheme, whereas
for EMLAM both schemes give similar predictions. It is also
noted that in the noise-free case, a smaller sampling time
$\tau_s$ improves the accuracy in the detection of the turning
points and consequently enhances the prediction with EMLAM,
whereas it perplexes the selection of the embedding parameters
$M$ and $\tau$ for LAM.

\begin{figure}[h!]
\centering
\centerline{\hbox{\includegraphics[height=60mm]{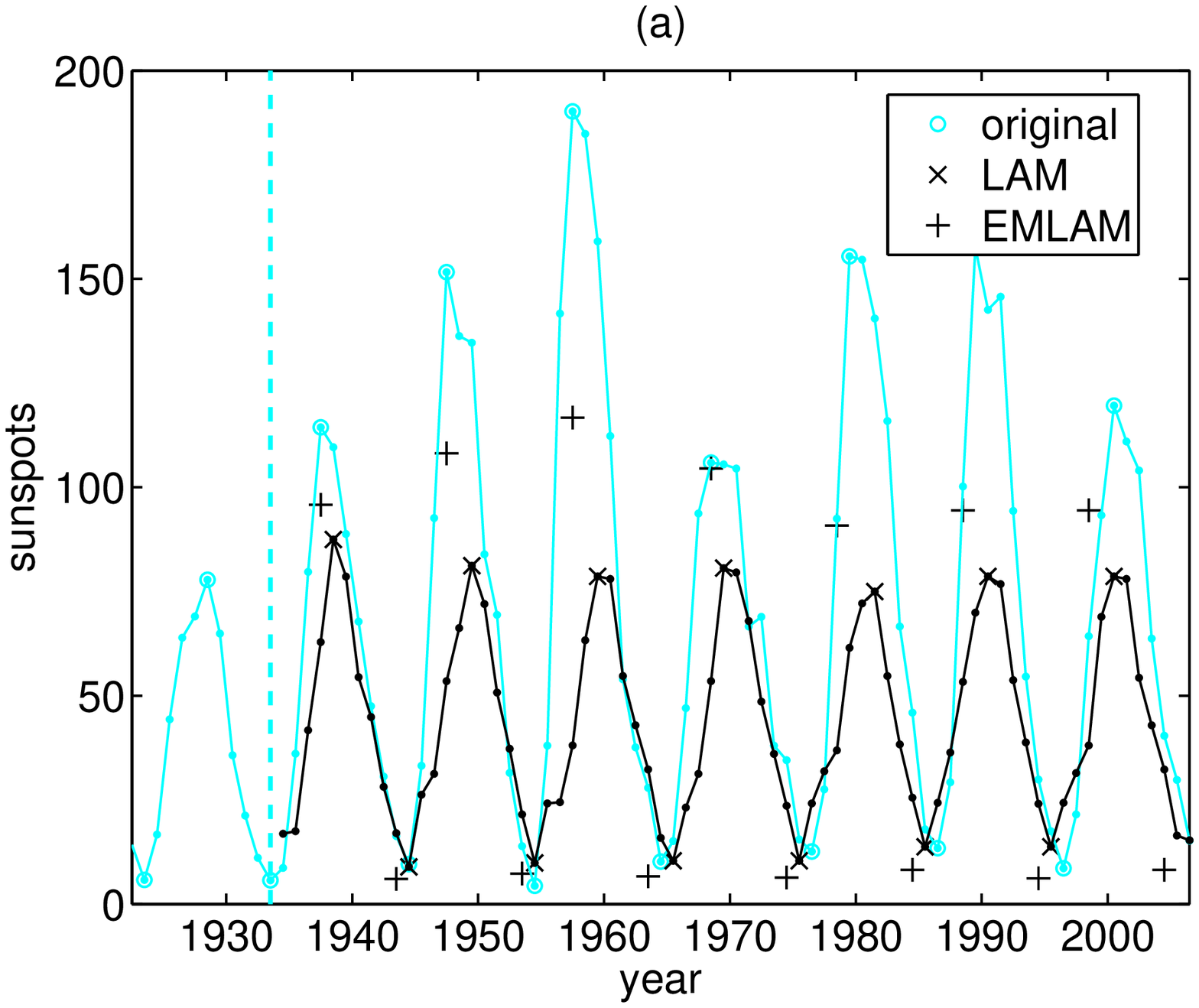}
\includegraphics[height=60mm]{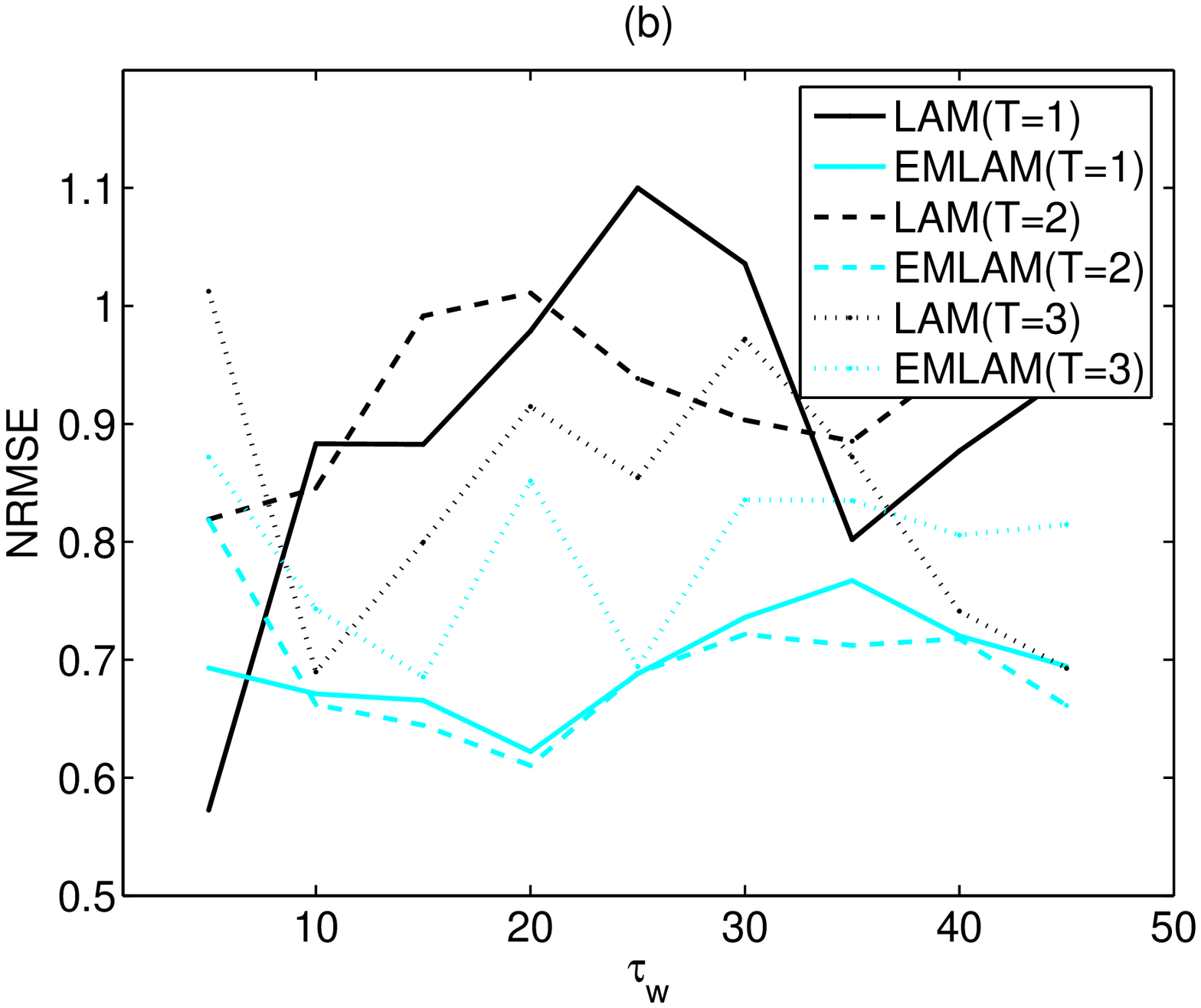}}}
\caption{Iterative prediction of turning points of sunspot numbers
with LAM and EMLAM. (a) Out-of-sample iterative prediction of the
last 70 annual sunspot numbers with LAM ($M=11$, $\tau=1$, $K=5$)
and EMLAM ($m=3$, $K=5$) as given in the legend. The turning points
were detected using $p=1$. The vertical line denotes the target
time. (b) NRMSE of iterative prediction of turning point magnitudes
at $T=1,2,3$ with LAM and EMLAM ($K=5$), as given in the legend,
computed on the last quarter of the sunspot time series, where
$\tau_w=M-1$ for LAM and $\tau_w=(m-1)5$ for EMLAM to account
roughly for the 11 year cycle.}
 \label{fig:predsunspots}
\end{figure}

We apply the same prediction setup to the celebrated time series of
annual sunspot numbers from year 1700 to 2006. Sunspot numbers
exhibit a rather regular oscillation of about 11 years long with
stable trough but varying peak that has given rise to debatable
prediction results suggesting stochastic, noisy periodic, and
chaotic behavior, obtained with time series models and other models,
such as models of input-output systems
\cite{Weigend94,Kugiumtzis98,Terui02,Pesnell07,Ukhorskiy02}. As
shown in Fig.~\ref{fig:predsunspots}a, the out-of-sample predictions
of LAM give rather periodic oscillations, failing to approximate the
true peaks, whereas EMLAM matches better the true peaks both in
magnitude and time.
The difference in LAM and EMLAM turning point prediction is rather
consistent over different embedding schemes, as shown in
Fig.~\ref{fig:predsunspots}b regarding the most favorable scenario
for LAM. Other state space reconstructions with $\tau>1$ as well as
the direct scheme gave worse LAM predictions. The summary results
in Table~\ref{tab:predsunspots} for different $K$ and $T$, show the
superiority of EMLAM over LAM, where again best EMLAM predictions
are obtained for small $m$.
\begin{table}[ht]
  \centering
   \begin{tabular}{|c|c|cc|cc|cc|cc|}
    \hline
    \multicolumn{2}{|c|}{} & \multicolumn{4}{c|}{Magnitudes} & \multicolumn{4}{c|}{Times}
    \\ \hline
    $T$ & $K$ & $M$ & \hspace{3mm} LAM & $m$ & EMLAM & $M$ & \hspace{3mm} LAM & $m$ & EMLAM \\ \hline
 1 & 1  & 41 & 0.548  & 4 & 0.480  & 11 & 0.874  & 10 & 0.778 \\
 1 & 5  & 6 & 0.661  & 5 & 0.622  & 11 & 0.795  & 10 & 0.595 \\
 1 & 10  & 31 & 0.763  & 10 & 0.714  & 6 & 0.743  & 5 & 0.550 \\
\hline
 2 & 1  & 31 & 0.735  & 4 & 0.433  & 11 & 1.322  & 10 & 1.597 \\
 2 & 5  & 41 & 0.747  & 4 & 0.601  & 6 & 1.451  & 2 & 0.940 \\
 2 & 10  & 31 & 0.776  & 8 & 0.681  & 26 & 1.569  & 4 & 1.066 \\
\hline
 3 & 1  & 6 & 0.568  & 5 & 0.793  & 41 & 2.020  & 8 & 2.185 \\
 3 & 5  & 31 & 0.824  & 4 & 0.761  & 6 & 1.960  & 4 & 1.189 \\
 3 & 10  & 31 & 0.801  & 2 & 0.761  & 26 & 1.732  & 3 & 1.385 \\
\hline
  \end{tabular}
  \caption{Summary results of NRMSE for
direct prediction of sunspot turning points structured as in Table~\ref{tab:MCHyperChaos},
where $M=6,11,\ldots,46$ ($\tau=1$) and $m=2,\ldots,10$.}
  \label{tab:predsunspots}
\end{table}

\begin{figure}[h!]
\centering
\centerline{\hbox{\includegraphics[height=60mm]{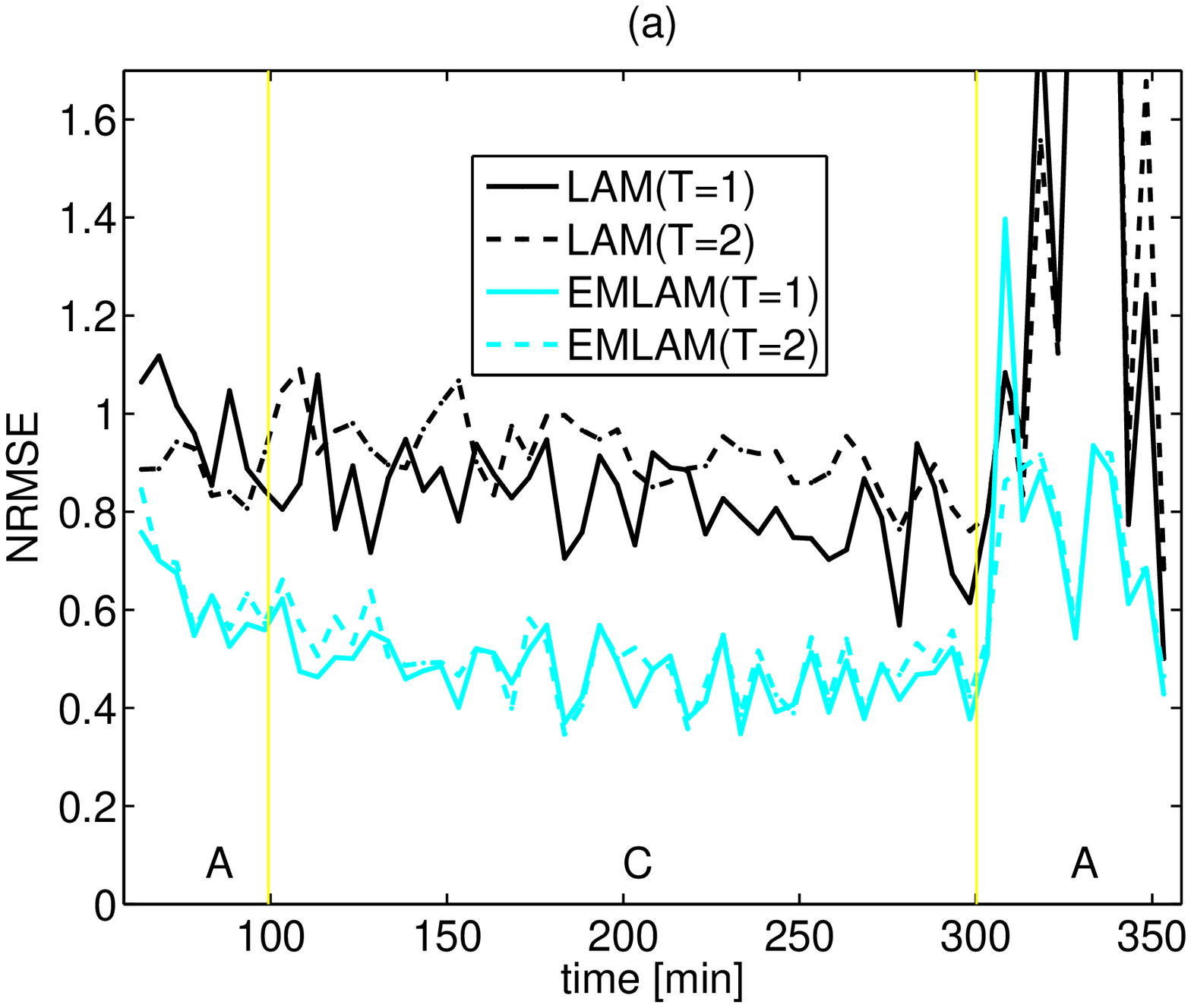}
\includegraphics[height=60mm]{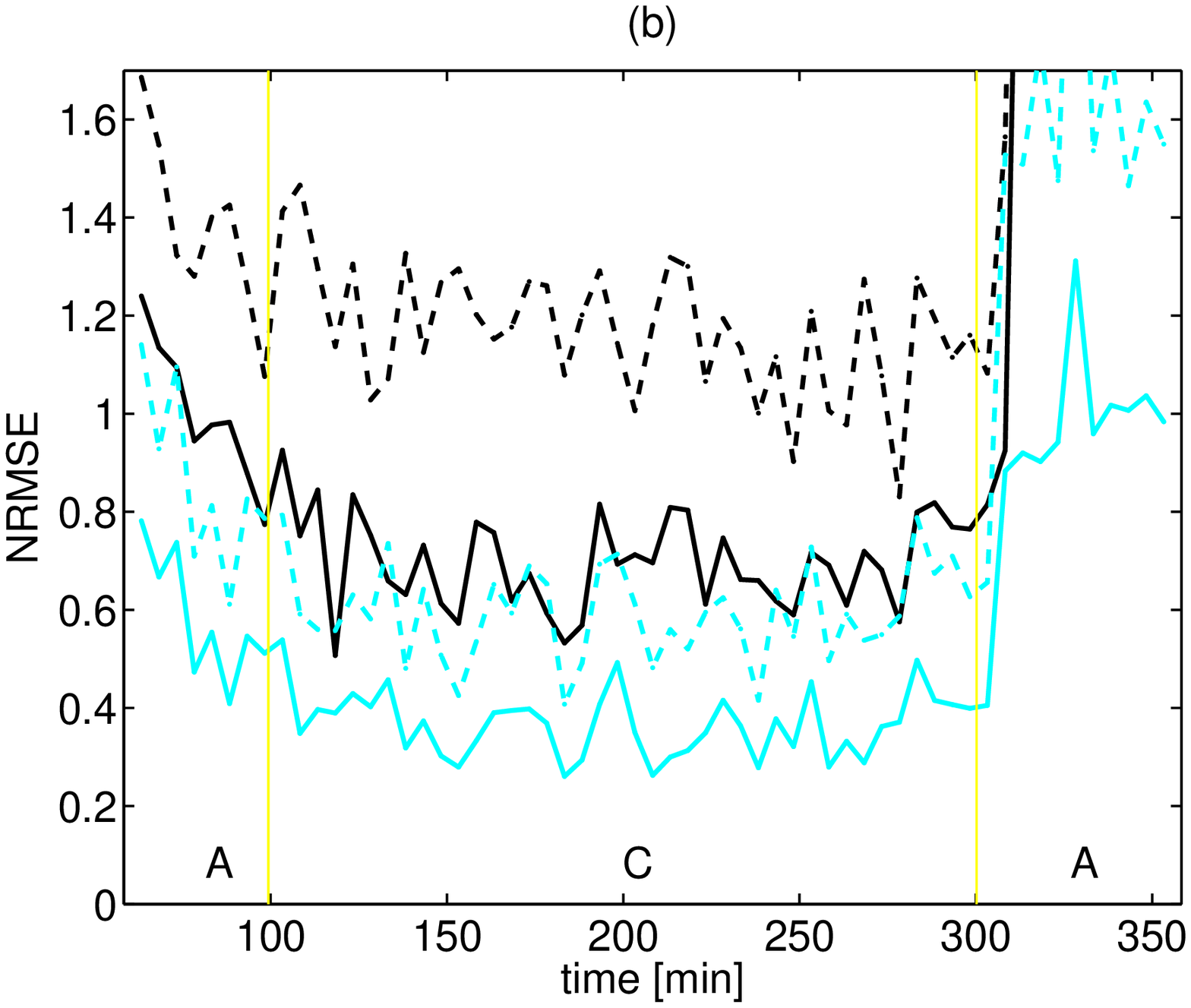}}}
\caption{Direct prediction of the magnitudes of turning points of
total stress in consecutive overlapping segments at times given in the abscissa.
The detection of
turning points is done with window of $p=1$ and zero-phase
filtering of order 13, and the model parameters are $K=10$,
$\tau=5$, $M=10$ and $m=5$. The predictions are for one and two time
step ahead for LAM and EMLAM as given in the legend.  The vertical grey lines
denote the separation of PLC band types denoted at the bottom. (b) The same
as in (a) but for the times of the turning points.}
 \label{fig:PLCprediction}
\end{figure}

Finally, we compare LAM and EMLAM on the time series of total stress
from an experiment of plastic deformation that exhibits the
{\em Portevin-Le Ch\^{a}telier (PLC) effect}. Poly-crystal Cu-15\%Al is tensile strained
at $\dot{\epsilon}=6.67\cdot10^{-6}\mbox{s} ^{-1}$ and
$T=125^{0}\mbox{C}$, and the total stress is sampled at $\tau_s=0.2\,\mbox{s}$; further details can
be found in \cite{Ziegenbein00,Kugiumtzis04a}. The increasing trend of total
stress was removed and the predictions were done on the last quarters of
overlapping segments of duration $1200\,\mbox{s}$ and sliding step $300\,\mbox{s}$.
For a range of model specific parameter values EMLAM gave consistently
better predictions than LAM. In Fig.~\ref{fig:PLCprediction}, the turning
point magnitude and time predictions with the direct scheme are shown for LAM and EMLAM
for arbitrary chosen model parameters.
EMLAM improves drastically the prediction of magnitudes and times of the
turning points for most of the epochs and the difference is larger for the
iterative prediction scheme (not shown here). LAM predictions are essentially at
the level of mean value prediction. To the contrary, EMLAM attains much
smaller NRMSE that varies in a way that allows the identification, at some degree,
of the transition of PLC band types: from type A to type C and then back to
type A, as shown with the vertical lines in Fig.~\ref{fig:PLCprediction}.
The actual transitions of PLC bands could only be identified by special
equipment following the local strains along the specimen
(for details see \cite{Ziegenbein00}). So, beyond improving the LAM prediction,
EMLAM prediction can possibly be used as a discriminating measure
for the PLC band types.
This point certainly bears further investigation.

In conclusion, this work suggests that the analysis of oscillating
time series, in particular those exhibiting rather linear upward and
downward trends, can be improved, and simplified, by restricting the
analysis to the turning points. It was shown that the information in
the time and magnitude of the turning points can be adequate to
explain the system dynamics. Simulations on a number of chaotic
flows and the two real-world examples showed that a local average
model based only on the turning points can predict turning points
equally to, or better than, the standard local average model,
a result of paramount importance for many applications.
There are a number of issues to be addressed,
such as implementation of other model types and inclusion of turning
point time in the model, but it seems that the focus on turning
points can give a new perspective in the analysis and prediction of
oscillating time series.


\end{document}